\documentclass[conference]{IEEEtran}
\IEEEoverridecommandlockouts
\usepackage{cite}
\usepackage{amsmath,amssymb,amsfonts}
\usepackage{algorithm} 
\usepackage{algorithmic}
\usepackage{graphicx}
\usepackage{textcomp}
\usepackage{xcolor}
\usepackage[a4paper, total={184mm,239mm}]{geometry}
\def\BibTeX{{\rm B\kern-.05em{\sc i\kern-.025em b}\kern-.08em
    T\kern-.1667em\lower.7ex\hbox{E}\kern-.125emX}}
\usepackage{fancyhdr} 
\renewcommand{\headrulewidth}{0pt} 

\begin{document}

	\fancypagestyle{plain}{ 
		\fancyhead{} 
		\fancyhead[L]{To appear in the proc. of the 35th Intern. IEEE Workshop on Rapid System Prototyping, 3rd Oct. 2024, Raleigh, NC, USA} 
		\renewcommand{\headrulewidth}{0pt} 
	}
	
\title{Transaction Level Hierarchy Guided and Functional Coverage Driven Deductive Formal Verification}

\author{\IEEEauthorblockN{Tobias Strauch}
	R\&D\\ EDAptix e.K.\\
	Munich, Germany\\
	Email: tobias@edaptix.com}

\maketitle
\maketitle \thispagestyle{plain}

\begin{abstract}
We demonstrate how dynamic verification (e.g. simulation) can be replaced by deductive formal verification and how to benefit from the advantages of symbolic verification and the reuse of verification proofs. To do this, we swap the well-known module-hierarchy based concept with a transaction-level (TL) based alternative, which still allows us to describe the design as precisely as on RTL. We enhance the aspect-oriented and TL oriented language PDVL to support the definition of functional coverage (FC) and assertions at all levels of a TL-hierarchy.

We then show how to use a deductive formal verification (DFV) flow which compiles PDVL code into Gallina code to be used by the Coq theorem prover. It can be argued that FC can be converted into proof obligations and that proving them is equivalent to 100\% coverage. We also demonstrate how lower-level proofs can be reused when verifying aspects at higher-levels of a TL-hierarchy. We argue that the traditional assertion-based verification (ABV) methodology is still supported and SVA can be proven using DFV.

\end{abstract}


\section{Introduction}

Verification can generally be divided into dynamic and static verification methods. Dynamic methods such as simulation are usually associated with a testbench (TB) that stimulates and monitors the design behavior. If the guidelines of the Portable Stimulus Specification (PSS) and the Universal Verification Methodology (UVM) are followed, then the functional coverage (FC) is defined and checked within testbenches. SystemVerilog Assertions (SVA) and their automatically derived coverage are typically defined within the Design Under Verification (DUV) and can be (dynamically) verified during simulation. Alternatively, static verification methods such as an SMT solver are used to statically verify SVA. 

We exploit the aspect-oriented and transaction-level (TL) language PDVL \cite{PDVL_spec}, which allows the definition of TL-hierarchies. We associate FC and assertions with virtual transactions on individual hierarchy levels. Then we convert the DUV and verification code into Gallina \cite{Gallina} code to be used for deductive formal verification (DFV). We argue that FC can be converted into theorems and that proving them is equivalent to 100\% FC.

In Section II FC as well as assertions and their coverage are discussed. Our work is described in Section \ref{work}. Then PDVL is reintroduced briefly and the conversion to Gallina code is outlined in Section \ref{MRPHS}. TL-hierarchy guided and functional coverage driven DFV is discussed in Section \ref{main}, followed by the presentation that the traditional SVA based ABV approach is also supported. The paper finishes with related work, results, and the concluding Section \ref{conclusion}.

\begin{figure*}[htbp]
	\centering
	\includegraphics[width=\textwidth,height=6.45cm]{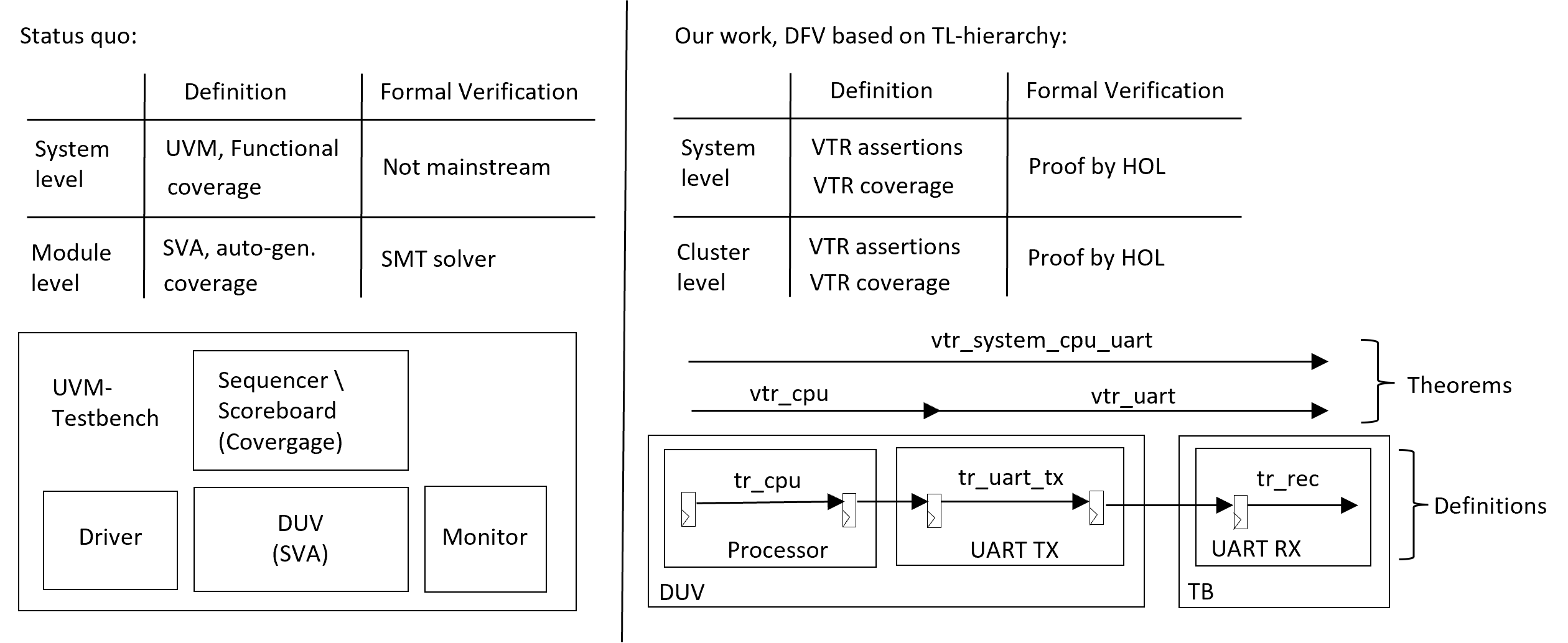}
	\caption{Comparing status quo and transaction-level-hierarchy guided and functional coverage driven deductive formal verification.}
	\label{fig_Concept}
\end{figure*}

\section{Functional coverage and assertions}
\label{assertions}

Today, FC comes in two flavors in SystemVerilog. One type of FC is sample-based coverage provided by a covergroup. Covergroups record the number of occurrences of various values specified as coverpoints. These coverpoints can be hierarchically referenced by testcases and testbenches so that it can be verified whether certain values or scenarios have occurred. They also provide a means for creating cross coverage. Unlike assertion-based cover properties, covergroups may be used in both class-based objects or structural code.

The second type of FC is assertion based. Assertions became very popular at a time when individual IP modules had to be connected to an SoC. Despite the introduction of an SoC module interconnect specification (e.g. AMBA from ARM), IP providers wanted to ensure that the bus interface of their encrypted IP block was functioning properly and was stimulated correctly. Since then, assertion-based verification (ABV) has gained greater acceptance and is now used for FC and the definition of lower-level assertions of various design aspects.

Temporal logic is widely used in property checking based formal verification as well as ABV. To express different types of properties, a variety of temporal logics have been proposed. For example, Linear-time Temporal Logic (LTL) is able to describe a property along with a single execution. However, it lacks the ability to express other possible executions and Computation Tree Logic (CTL) is proposed to solve the problem. 
The assertion definitions are used to (auto-)generate assertion-based FC points.

The relevant type of coverage comes from a cover property, which uses the same temporal syntax as defined by SVA. Since cover properties use the same properties as asserts, the same work in creating the properties can be reused in both checking and coverage gathering. Cover properties are typically used for protocol coverage since the temporal syntax is ideal for describing sequences of events over time, such as those required for bus interfaces.

However, cover properties can only be placed in structural code (i.e., modules, programs, or interfaces) and cannot be used in class-based objects. Likewise, their coverage information is not easily accessible in SystemVerilog (SV) for use in a testbench (e.g. steering stimulus generation).

Fig. \ref{fig_Concept} shows that SVAs are defined at the module level according to the current state of the art. Dynamic ABV can be based on simulation, emulation or FPGA based prototyping. All of them face the challenge of activating an assertion. When dynamic ABV is combined with constraint random or directed verification, the runtime can be very high to activate coverage points of corner cases. HW-based dynamic ABV solutions have the problem of collecting relevant assertion coverage data. Static ABV uses predominantly SMT solver to verify the correctness of assertions.

\section{Our work}
\label{work}

Our work enables the definition of FC and assertions from the lower-level up to the system-level and supports their coverage-driven DFV. It is set in contrast to the status quo in Fig. \ref{fig_Concept}.

\subsubsection{Adding relevant language constructs to PDVL} PDVL is an aspect-oriented and transaction-level Programming Design and Verification Language, which itself adds language constructs on top of SystemVerilog (SV). 

In our work we demonstrate further improvements to PDVL to enable DFV from lower-level up to the system-level. We refer to code that asserts expected behavior as virtual transactions (VTRs) in PDVL. VTRs are not meant to be synthesizable unless explicitly specified.

\subsubsection{Defining FC and assertions from lower-level up to system-level} When using SV, system-level coverage points can be defined to capture complex system behavior. To make optimal use of this possibility and to support a certain level of reuse, the UVM and the PSS were defined. Using SVA at the system level to assert complex system behavior remains challenging. 

In our work we demonstrate the use of VTRs that can bundle a set of less complex VTRs. This allows the definition of a hierarchy of VTRs which asserts behavior from system-level all the way down to lower-level. Higher-level VTRs can therefore also assert system behavior that spans over DUV and TB behavior alike. This reusable TL-hierarchy based on VTRs seamlessly fills the gap which exists between lower-level SVA and the expected behavior defined by the UVM.

\subsubsection{Verifying FC and assertions from lower-level up to system-level} System-level coverage is primarily captured using simulation techniques. As symbolic simulation is still limited, many simulation runs must be executed by directed or constraint random tests. Additionally, the possibility to reuse verification throughout the module hierarchy is limited. Verifying system-level coverage using static methods such as DFV continues to pose significant challenges, and its usage cannot be considered mainstream today.

In our work we enable VTRs to define coverage. VTRs can be used to assert behavior seamlessly throughout the complete hierarchy, from lower-level up to the system-level. Therefore, coverage can be defined and verified throughout the TL-hierarchy as well. We demonstrate how these VTRs and their associated coverage can be converted into Gallina code and how the individual coverage points can be proven. We argue that individual coverage points of VTRs can be associated with individual proofs. Subsequently, VTRs which are built on other VTRs can define coverage that can be proven by reusing proofs of their lower-level VTRs.

\subsubsection{Using lower-level assertion definition and associated coverage} Lower-level assertion definitions (such as SVA) are very well established for lower-level logic such as finite state machines (FSMs) and interface protocols. Usually, assertion coverage definition is automatically derived from these lower-level assertions and can be verified statically by SMT solvers or during dynamic verification (e.g. simulation). 

Our work demonstrates how to reuse SVA in PDVL and discusses the conversion of the design and the lower-level assertion definitions into Gallina code. Combined with the compiled design behavior, the assertions can then be verified by proving using a DFV tool like the Coq proof assistant \cite{Coq}.  

\subsubsection{Proof by symbolic simulation}

In our work we focus on DFV based on symbolic simulation, as demonstrated in \cite{SEC}. Theorems can be proven that for a given initial state and a given stimulation sequence, where state and sequence values can have a symbolic type, all possible resulting states and state transitions match the expected behavior.

\section{PDVL}
\label{PDVL}

PDVL was introduced in 2017 \cite{PDVL}. We therefore only give a very brief overview. PDVL is based on SV and encapsulates conditions and assignments. The latter are then called datapaths. Transactions (TRs) determine the conditions under which individual datapaths are valid. PDVL is not limited to CPUs, but an example of a RISC-V instruction is given in Alg. \ref{alg_riscv_instruction}. 

Aforementioned elements are stored in clusters. The aspect-oriented paradigm of PDVL becomes obvious, when looking at the hardware generation process (Alg. \ref{alg_pdvl_joining}). A hardware module hierarchy is built, and clusters are then joined into the individual modules. The merging and signal routing must be handled by the compiler. More information can be found in the PDVL specification \cite{PDVL_spec}.

In this paper, we introduce the concept of VTRs which allows us to define a TL-hierarchy (Fig. \ref{fig_TLHierary}). We also show by examples, which SV constructs are reused in PDVL for FC and assertion definition.

\section{PDVL to Gallina compiler}
\label{MRPHS}

The tool “MRPHS” compiles PDVL code into synthesizable SV code. MRPHS’ PDVL to Gallina compiler extension was introduced in \cite{PDVL_Gallina}. We therefore only give a very brief overview.



The design “DUV” must be generated by using the relevant PDVL build commands (Alg. \ref{alg_pdvl_joining}) before PDVL code is compiled into a Gallina representation (Alg. \ref{alg_coq_basic}). Once the logic is joined, all sequential and combinatorial signals of the DUV are identified.  


We mention the standard Boolean type "bool" for signal bits in this paper, but any user-defined type can be used instead. Alg. \ref{alg_coq_basic} shows that all signals are added to an inductively defined type t\_item and that a dynamic list t\_state can be generated based on t\_item members. 

Only signals with a defined value are added to the t\_state list. A signal can be removed from the list once its value becomes undefined. This applies to both sequential and combinatorial signals. The t\_state list therefore contains all sequential elements and combinatorial signals that are defined at a given point of time.

Alg. \ref{alg_coq_basic}, line 12 shows that design-specific functions are defined, which are subsequently used in the generated Gallina code. MRPHS also generates a set of more general functions which, for instance, assign a given value to an item in the t\_state list (e.g. f\_set) or extract the value of a specific item in the t\_state list (e.g. f\_get). The generated Gallina code mentioned so far is stored as a DUV specific library.


For each condition in the PDVL source code, a Boolean signal is generated and added to the item list as well. The condition body is converted into a Gallina definition.

All condition, datapath, and transaction definitions in PDVL are compiled into Gallina code and stored as a second DUV-specific library, which is used by the final proof scripts. Alg. \ref{alg_gallina_design} gives an example related to the aforementioned RISC-V ADDI instruction. We will see in the following section, that theorems can be proven based on the automatically generated Gallina code mentioned so far.






\begin{algorithm}
	\caption{PDVL: RISC-V instruction example (ADDI)}
	\begin{algorithmic}[1]
		\STATE \textcolor{blue}{cl\_}instr\_addi \{\space \space \space \space \space \space \space \space \space \space \space \space \space \space \space \space \space \space \space \space \space \space \space \space \space \space \space \space \space \space \space \space \space \space \space \space \space (* cluster *)
		\STATE \textcolor{blue}{c\_}instr\_i\_addi \{ if (opcode\_i == 7'h13\space \space \space \space \space  (* condition *)
		\STATE $ $ \space \& funct3i == 3'h0) this; \}
		\STATE \textcolor{blue}{d\_}addi \{ dp\_out = rs1\_dato+instr[31:20]; \} \space (* datapath *)
		\STATE \textcolor{blue}{tr\_}rv32i\_addi \{ \space \space \space \space \space \space \space \space \space \space \space \space \space \space \space \space \space \space \space \space \space \space \space \space \space \space \space \space \space \space \space \space \space (* transaction *)
		\STATE $ $ \space unique @c\_instr\_i\_addi \{ d\_rs1i\_addr; d\_addi; 
		\STATE $ $ \space $ $ \space d\_rd\_dp\_out; d\_rd\_addr; c\_rf\_write;  d\_pc4; \} \} \}
	\end{algorithmic}
	\label{alg_riscv_instruction}
\end{algorithm}

\begin{algorithm}
	\caption{PDVL: Generating a module hierarchy, joining a cluster into a submodule and defining clock input and edge sensitivity for registers.}
	\begin{algorithmic}[1]
		\STATE \textcolor{blue}{build} TB \{
		\STATE $ $ \space \textcolor{blue}{build} i\_duv DUV; \space \space \space \space \space \space \space \space \space \space \space \space \space \space \space \space \space \space \space \space (* build hierarchy *)
		\STATE $ $ \space \textcolor{blue}{join} cl\_rv32i cl\_rv32imc; \space \space \space \space \space \space \space \space \space \space \space \space \space \space \space \space \space \space \space \space \space \space \space (* joining *)
		\STATE $ $ \space \textcolor{blue}{join} \{ tr\_reg \{ @e\_clk \{ tr\_rv32i\_addi; \}\}\} cl\_rv32imc;
		\STATE $ $ \space \textcolor{blue}{join} cl\_rv32imc i\_duv; \}
	\end{algorithmic}
	\label{alg_pdvl_joining}
\end{algorithm}

\begin{algorithm}
	\caption{Gallina: Signal definition examples, the design state list and a boolean function example}
	\begin{algorithmic}[1]
		\STATE \textcolor{blue}{Inductive} t\_item : Type := 
		\STATE $ $ \space $|$ i\_nil 
		\STATE $ $ \space $|$ instr ( l : t\_bus32 )
		\STATE $ $ \space $|$  ...
		\STATE $ $ \space $|$ pc ( l : t\_bus20 )
		\STATE $ $ \space $|$ reg\_file ( l : t\_arr32x32 ) .
		\STATE $ $ \space
		\STATE \textcolor{blue}{Inductive} t\_state : Type := 
		\STATE $ $ \space $|$ st\_nil 
		\STATE $ $ \space $|$ st\_cons ( s : t\_item ) ( l : t\_state ) .
		\STATE $ $ \space
		\STATE \textcolor{blue}{Definition} f\_equal32 ( a b : t\_bus32 ) : bool :=  ...
	\end{algorithmic}
	\label{alg_coq_basic}
\end{algorithm}

\begin{algorithm}
	\caption{Gallina: Compiled condition, datapath and transaction definitions (ADDI example)}
	\begin{algorithmic}[1]
		\STATE \textcolor{blue}{Definition} c\_instr\_i\_addi ( st : t\_state ) : t\_state := ...
		\STATE \textcolor{blue}{Definition} d\_addi ( st : t\_state ) : t\_state := ...
		\STATE \textcolor{blue}{Definition} tr\_rv32i\_addi ( st : t\_state ) : t\_state := ...
	\end{algorithmic}
	\label{alg_gallina_design}
\end{algorithm}

MRPHS generates Gallina definitions to support symbolic simulation techniques. The definition sim\_update (Alg. \ref{alg_gallina_sim}, line 1) walks through the complete design and updates all possible non-sequential signals. Assuming the DUV has only one single clock, then the definition sim\_cycle (Alg. \ref{alg_gallina_sim}, line 1) simulates a complete cycle, which includes an update of all sequential elements, followed by an update of all non-sequential signals. For more complex clock domain structures, the sim\_cycle definition is adapted accordingly. Alg. \ref{alg_gallina_sim}, line 3 demonstrates, how two cycles of the design can be simulated. 

\begin{algorithm}
	\caption{Gallina: Helper definitions for symbolic simulation}
	\begin{algorithmic}[1]
		\STATE \textcolor{blue}{Definition} sim\_update ( st : t\_state ) : t\_state := ...
		\STATE \textcolor{blue}{Definition} sim\_cycle ( st : t\_state ) : t\_state := ...
		\STATE \textcolor{blue}{Definition} two\_cycles ( st : t\_state ) : t\_state := 
		\STATE $ $ \space sim\_cycle (sim\_cycle st).
	\end{algorithmic}
	\label{alg_gallina_sim}
\end{algorithm}


\section{TL-hierarchy guided coverage driven DFV}
\label{main}

\subsection{Introducing virtual transactions} 

In this section we introduce virtual transactions (VTRs) which allow us to define a TL-hierarchy and ultimately the reuse of lower-level verification results on higher-level. VTRs can define testbench related functionality such as sequences, randomness, coverage points, etc.. A VTR can bundle a set of TRs and VTRs. This generates a TL-hierarchy that can have multiple top-level VTRs, whereas each of the top-level VTRs groups a specific set of TRs and VTRs for a specific verification goal. VTRs are not meant to be synthesizable unless explicitly specified.

One of the key ideas behind PDVL is that it extends the SV language by only a limited number of new constructs. It is therefore intended that VTRs in PDVL reuse many of the testbench related language constructs known from SV. We mention some keywords by providing examples.

\subsection{VTRs and testbench related language constructs} 

\begin{figure*}[htbp]
	\centering
	\includegraphics[width=\textwidth,height=7.0cm]{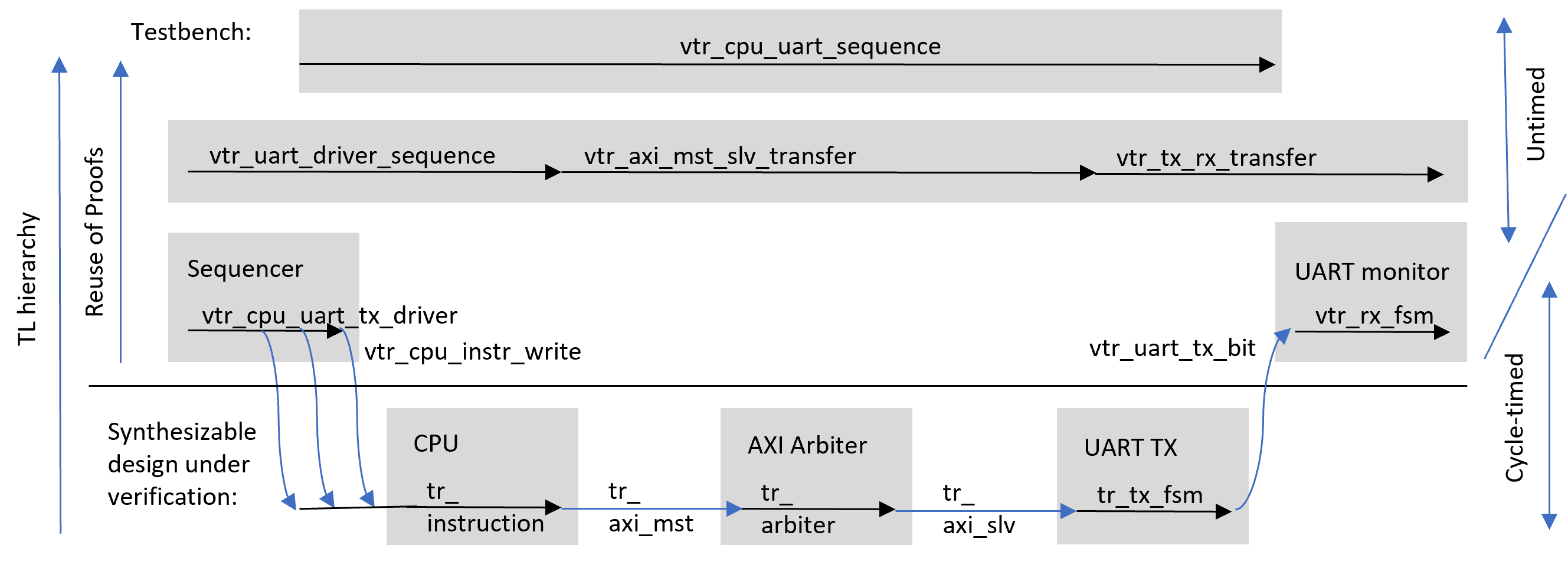}
	\caption{Transaction level hierarchy example, showing the design under verification (DUV) and transactions as well as the testbench (TB) and virtual transactions.}
	\label{fig_TLHierary}
\end{figure*}

Fig. \ref{fig_TLHierary} gives an overview of the design example we use to demonstrate our methodology. The synthesizable UART transmit (TX) module in the DUV provides standard UART TX capabilities. The outgoing datastream is captured by an UART monitor. We now show how a VTR (Alg. \ref{alg_vtr_trans_uart}) can be defined, which creates a verification environment for these two entities.

A VTR can define time consuming behavior by using the keyword sequence (Alg. \ref{alg_vtr_trans_uart}, line 5). A sequence always starts at the “init” state, can traverse a finite number of user defined states (here “finish”) and becomes redundant when reaching the keyword “exit” (Alg. \ref{alg_vtr_trans_uart}, line 13).  A sequence can therefore also be considered as an FSM.

In our example (Alg. \ref{alg_vtr_trans_uart}, line 6), a random “symbolic” value is assigned to the byte “axi\_tx\_data” during the “init” state and the condition “c\_axi\_trans” is set valid. The sequence then transitions to the finish state. Once the condition “c\_uart\_rx\_valid” is true, it is checked whether the coverage point “cp\_tx\_rx\_eq” is covered and the sequence is exited.

If the VTR “vtr\_tx\_rx\_transfer” is called within a VTR that defines the clock for the registers involved, then the sequence is clocked by the specified clock (e.g. UART clock). In this case we are talking about a cycle-timed VTR.

A cycle-timed VTR checks signal values and condition states in the respective cycle. It also assigns condition states and signal values to the relevant signal at the given clock edge.

\begin{algorithm}
	\caption{PDVL: VTR UART TX-RX Transfer}
	\begin{algorithmic}[1]
		\STATE \textcolor{blue}{cluster} tb\_axi\_uart \{
		\STATE $ $ \space \space \textcolor{blue}{c\_}tx\_rx\_data\_eq \{ 
		\STATE $ $ \space \space \space \space \space if (uart\_rx\_data == axi\_tx\_data) this; \}
		\STATE $ $ \space \space \textcolor{blue}{vtr\_}tx\_rx\_transfer \{
		\STATE $ $ \space \space \space \space \space \textcolor{blue}{sequence} tx\_rx\_transfer \{
		\STATE $ $ \space \space \space \space \space \space \space \space \textcolor{blue}{init}: \{ 
		\STATE $ $ \space \space \space \space \space \space \space \space \space \space \space \textcolor{blue}{random} axi\_tx\_data;
		\STATE $ $ \space \space \space \space \space \space \space \space \space \space \space c\_axi\_trans;
		\STATE $ $ \space \space \space \space \space \space \space \space \space \space \space finish; \}
		\STATE $ $ \space \space \space \space \space \space \space \space finish: \{
		\STATE $ $ \space \space \space \space \space \space \space \space \space \space \space @c\_uart\_rx\_valid \{
		\STATE $ $ \space \space \space \space \space \space \space \space \space \space \space \space \space \space \space \space \textcolor{blue}{cover cp\_}tx\_rx\_eq \{ c\_rx\_tx\_data\_eq; \} 
		\STATE $ $ \space \space \space \space \space \space \space \space \space \space \space \space \space \space \space \space \textcolor{blue}{exit}; \} \} \} \} \}
	\end{algorithmic}
	\label{alg_vtr_trans_uart}
\end{algorithm}

\subsection{VTRs as instruction code generator}

A VTR can call other VTRs. The calling VTR can then be considered as an untimed VTR, which has more the character of a function written in a sequential programming language. The called VTR starts at the “init” state of the sequence. The called VTR finishes when the “exit” keyword is reached. At this timepoint, the calling VTR continues processing its sequence. 

The execution of the resulting VTR tree becomes time consuming when a VTR calls a cycle-timed VTR. Our enhanced PDVL version supports the fork-join and other known mechanisms to run multiple time-consuming sequences in parallel.

An example of an untimed VTR is an instruction generator for an UART TX driver shown in Alg. \ref{alg_vtr_trans_cpu} called “vtr\_cpu\_uart\_tx\_driver”. It is also outlined in Fig. \ref{fig_TLHierary} in the sequencer block, indicating that it calls the CPU multiple times to force the CPU to issue individual bus master writes.

After assigning a symbolic value to “axi\_tx\_data”, the VTR successively calls a list of VTRs. The goal is that the UART TX Enable Bit is set (“vtr\_cpu\_uart\_tx\_enable”) and that the “axi\_tx\_data” value is written into the UART TX register (“vtr\_cpu\_uart\_tx\_data”) which starts the transmission. Finally, the CPU should constantly loop over a NOP instruction (“vtr\_cpu\_loop\_nop”). The called VTRs force the CPU in the given system to execute individual instructions, which results in individual AXI bus-writes. 

The VTR “vtr\_cpu\_uart\_tx\_data” (Alg. \ref{alg_vtr_trans_cpu}, line 10) shows how the CPU can be forced to execute a sequence of instructions to issue the relevant AXI master data write command.

VTRs can be untimed, timed and cycle-timed. When a VTR is called by a VTR outside an edge sensitive block, the VTR is untimed and the VTRs interact by an calling-init-exit-continue process. Multiple VTRs can also interact using a fork-join or alternative mechanisms.

\begin{algorithm}
	\caption{PDVL: CPU UART Driver Instruction Generator}
	\begin{algorithmic}[1]
		\STATE \textcolor{blue}{cluster} tb\_cpu\_uart\_driver \{
		\STATE $ $ \space \space \textcolor{blue}{vtr\_}cpu\_uart\_tx\_driver \{
		\STATE $ $ \space \space \space \space \space \textcolor{blue}{sequence} cpu\_uart\_driver \{
		\STATE $ $ \space \space \space \space \space \space \space \space \textcolor{blue}{init}: \{ 
		\STATE $ $ \space \space \space \space \space \space \space \space \space \space \space \textcolor{blue}{random} axi\_tx\_data;
		\STATE $ $ \space \space \space \space \space \space \space \space \space \space \space \textcolor{blue}{vtr\_}cpu\_uart\_tx\_enable;
		\STATE $ $ \space \space \space \space \space \space \space \space \space \space \space \textcolor{blue}{vtr\_}cpu\_uart\_tx\_data;
		\STATE $ $ \space \space \space \space \space \space \space \space \space \space \space \textcolor{blue}{vtr\_}cpu\_loop\_nop;
		\STATE $ $ \space \space \space \space \space \space \space \space \space \space \space \textcolor{blue}{exit}; \} \} \} 
		\STATE $ $ \space \space \textcolor{blue}{vtr\_}cpu\_uart\_tx\_enable \{ ... \}
		\STATE $ $ \space \space \textcolor{blue}{vtr\_}cpu\_uart\_tx\_data \{
		\STATE $ $ \space \space \space \space \space \textcolor{blue}{sequence} cpu\_tx\_data \{
		\STATE $ $ \space \space \space \space \space \space \space \space \textcolor{blue}{init}: \{ instr2;
		\STATE $ $ \space \space \space \space \space \space \space \space \space \space \space ... /* code executing "lui s0, 0x80030;" */ \}
		\STATE $ $ \space \space \space \space \space \space \space \space 		\textcolor{blue}{instr2}: \{ instr3;
		\STATE $ $ \space \space \space \space \space \space \space \space \space \space \space ... /* code executing "li a5, axi\_tx\_data;" */ \}\\
		\STATE $ $ \space \space \space \space \space \space \space \space 		\textcolor{blue}{instr3}: \{
		\STATE $ $ \space \space \space \space \space \space \space \space \space \space \space ... /* code executing "sw a5, 4(s0);" */ \}\\
		\STATE $ $ \space \space \space \space \space \space \space \space \space \space \space \textcolor{blue}{exit}; \} \} \}
		\STATE $ $ \space \space \textcolor{blue}{vtr\_}cpu\_loop\_nop \{ ... \}\}
	\end{algorithmic}
	\label{alg_vtr_trans_cpu}
\end{algorithm}

\subsection{VTRs and coverage related constructs}

PDVL supports the concept of SV to assign a random value to a signal. This is the case, for example, if the payload of a data transmission is generic. PDVL uses the keyword “random” to indicate, that any valid value within the given range of a signal must be considered during verification (Alg. \ref{alg_vtr_trans_uart}, line 7 and Alg. \ref{alg_vtr_trans_cpu}, line 5). In the remainder of the paper we use the term "symbolic" value for such an assignment.

We also introduce a coverage definition for individual VTRs. A coverage point is defined by the keyword “cover”, followed by a property name and a body that, for example, checks whether a certain condition is true at a given cycle (Alg. \ref{alg_vtr_trans_uart}, line 12).


\subsection{TL-hierarchy} 

So far we have discussed the CPU and the UART section of our example design. We need to mention the AXI arbiter block, which basically converts AXI master writes to the relevant AXI slave writes. A VTR can also be defined for an AXI master-slave transfer, such as “”vtr\_axi\_mst\_slv\_transfer” in Fig. \ref{fig_TLHierary}.

There are additional TRs and VTRs which connect the design modules and testbench elements. Fig. \ref{fig_TLHierary} illustrates that all TRs are part of the synthesizable design. We have also discussed VTRs which give an abstract view of local behavior such as a CPU sequence and a UART TX-RX data transfer. 

We now broaden the perspective and introduce a VTR which produces a system-wide sequence, shown in Fig. \ref{fig_TLHierary}. In our example, the top-level sequence “vtr\_cpu\_uart\_sequence” forces the CPU to generate a set of instructions which enable the UART to transmit data. The CPU then writes a symbolic value into the UART, which is then transmitted and captured by the UART monitor. Fig. \ref{fig_TLHierary} illustrates that it can be seen as a top-level VTR covering previously mentioned VTRs so far. We will see how our approach to use DFV will benefit from such a TL-hierarchy.



\subsection{VTRs and deductive formal verification} 

We now show how VTRs and the resulting TL-hierarchy can be used for DFV. We improved the PDVL to Gallina compiler (“MRPHS”, Section \ref{MRPHS}) to support VTRs.

\subsubsection{Proving theorems based on combined transactions}

Alg. \ref{alg_VTRtheorem} shows a simplified code of the example VTR (Alg. \ref{alg_vtr_trans_uart}) compiled into Gallina code. The theorem lists a symbolic value as a hypothesis. Coverage properties are compiled into definitions, which check that the defined conditions are true (“cp\_uart\_tx\_rx”). The sequence “vtr\_tx\_rx\_transfer” is also compiled into a definition, which modifies the list of the given design states. The compiled and provable theorem “th\_tx\_rx\_transfer” executes the “vtr\_tx\_rx\_transfer” and checks whether the property “cp\_uart\_tx\_rx” is covered.

\subsubsection{Proving theorems based on sequences}

In this sub-section, we refer to the CPU UART driver code generator described in Alg. \ref{alg_vtr_trans_cpu}. To prove that each of the code generation sequences (vtr\_cpu\_uart\_tx\_enable and vtr\_cpu\_uart\_tx\_data) initiates correct AXI master writes, individual cover properties can be defined similar to the coverage property in Alg. \ref{alg_vtr_trans_uart}.

\begin{algorithm}
	\caption{Gallina: Coverage theorem (simplified)}
	\begin{algorithmic}[1]
		\STATE \textcolor{blue}{Theorem} th\_tx\_rx\_transfer : 
		\STATE  $ $ \space \space \space foreach axi\_tx\_data : t\_bus, 
		\STATE  $ $ \space \space \space cp\_uart\_tx\_rx (
		\STATE  $ $ \space \space \space vtr\_tx\_rx\_transfer ( reset\_st )).
	\end{algorithmic}
	\label{alg_VTRtheorem}
\end{algorithm}

\begin{algorithm}
	\caption{Gallina: CPU UART driver sequence}
	\begin{algorithmic}[1]
		\STATE \textcolor{blue}{Theorem} th\_cpu\_uart\_driver : 
		\STATE  $ $ \space \space \space foreach axi\_tx\_data : t\_bus, 
		\STATE  $ $ \space \space \space cp\_cpu\_tx\_data (
		\STATE  $ $ \space \space \space cp\_cpu\_tx\_enable (
		\STATE  $ $ \space \space \space vtr\_cpu\_loop\_nop (
		\STATE  $ $ \space \space \space vtr\_cpu\_uart\_tx\_data ( axi\_tx\_data
		\STATE  $ $ \space \space \space vtr\_cpu\_uart\_tx\_enable ( reset\_st ))))).
	\end{algorithmic}
	\label{alg_gallina_sequence}
\end{algorithm}

Alg. \ref{alg_gallina_sequence} lists the compiled and provable theorem for the “cpu\_uart\_driver” sequence listed in Alg. \ref{alg_vtr_trans_cpu}. After the individual sequences have been executed, it is checked whether the relevant properties are covered (“true”).

\subsection{Reuse of proofs}

The ability to define more abstract VTRs to build a TL-hierarchy can be particularly useful when proofs of lower-level theorems derived from lower-level TRs or VTRs can be reused. We start with a top-level VTR.

\subsubsection{Top-level VTR reuses lower level proofs}

We can see in Fig. \ref{fig_TLHierary} how the top-level VTR “vtr\_cpu\_uart\_sequence” functionally extends over all VTRs mentioned so far. We want to prove that the symbolic value which is programmed by the CPU is transmitted from the UART TX module to the UART monitor. This can be asserted within a VTR similar to the “cp\_uart\_tx\_rx” cover property in Alg. \ref{alg_vtr_trans_uart}, line 12. The top-level VTR “vtr\_cpu\_uart\_sequence” needs to call the same sequence as defined in Alg. \ref{alg_vtr_trans_cpu}, lines 6-8.  

After compiling the VTR “vtr\_cpu\_uart\_sequence” to the relevant theorem, the proof can reuse proven theorems of lower-level theorems such as the one of “th\_tx\_rx\_transfer” etc..

\subsubsection{Mid-level VTR reuses lower level proofs}

When proving mid-level theorems such as “th\_tx\_rx\_transfer”, reusing lower-level theorems (not previously mentioned) is also advantageous.

In the given UART example, it can be argued that a finite state machine (FSM) controlling the UART TX has its counterpart in the FSM of the UART receiver (RX) in the testbench monitor. For simplicity, let's assume that the transmission is based on a synchronous single-bit protocol.

The TR in the TX-FSM that defines the transmission of a single bit is combined with the TR in the RX-FSM to form a new VTR that defines the entire transmission process of a single bit value. This abstract VTR in PDVL is compiled into Gallina code which can be used for proving. A theorem is proven stating that the transmitted data bit value is correct.

The same mechanism is applied for the transfer of a byte and a full packet, while using the previously formally verified theorem of transmitting a bit (or byte respectively).


\subsection{SVA in PDVL and intermediate representation}
The work in \cite{Syn_SVA} shows how SVAs can be converted into synthesizable code by extracting FSMs, datapath and checker logic. We follow these guidelines and generate an IR that is in-line with PDVL representations. Local SVA variables are converted into PDVL items. The Boolean layer is compiled into condition and datapath logic as defined by PDVL. SVA sequences and properties are converted into VTR sequences as defined in this paper. Here multiple VTR sequences can result from this conversion process and the possibility to use fork-join methods becomes relevant. SVA local variables can also be compiled into registers within the IR. Assumptions and coverage goals are converted into coverage properties. 

At the current state, our work does not support the full range of SVAs. There are limitations in aspects like liveliness, overlapping transactions, etc.. However, it is not our goal to fully support this specific methodology. Let’s assume there are SVAs in a sender and a receiver of an interface. We argue that our approach of a TL-hierarchy covers both sides with VTRs spanning over sender and receiver functionality, so that local SVAs become redundant.





\section{Related work}

BSV: Bluespec SystemVerilog (BSV) \cite{Bluespec} is a high-level hardware description language of guarded atomic actions. The BSV concept is based on BSV rules and a term rewriting system, whereas each rule can be viewed as a declarative assertion expressing a potential atomic state transition. 
In \cite{BSV_MOD} it is discussed, how the modular concept of BSV generates new challenges for predicting the compiler output. 
The sequential behavior of the hardware defined by PDVL is exact and the cycle behavior of the compiler’s SV and Gallina code is therefore predictable since no scheduling is involved. 

KAMI: According to \cite{KAMI}, KAMI is a framework to support implementing, specifying, formally verifying, and compiling hardware designs based on BSV and the Coq theorem prover. It emphasizes modular verification of digital hardware. In contrast, using PDVL provides a TL design paradigm and a TL formal verification paradigm. TRs can span multiple modules, creating a TL-hierarchy that can range from low-level cycle-accurate TRs to approximately-timed or untimed VTRs. The concept of a module becomes only relevant when the final SV code is generated.



Rules rewriting: The work in \cite{PSL_prover} describes the proving and disproving of assertion rewrite rules with automated theorem provers. The work is based on the assertion language PSL and concentrates on rule rewriting. We outlined our flow to convert SVA into an IR of FSMs, datapath and checker logic, which is then compiled into Gallina code and used for DFV. We do not use rewriting as such, but benefit greatly from reusing proofs through applying proven theorems.


RTL to TLM: The main intent of the work in \cite{RTL2TLM} is to automatically build a dynamic ABV environment for a TLM model, with no restrictions on the abstraction level, by starting from a set of properties initially defined for a corresponding RTL implementation. First, cycle-accurate RTL properties are automatically rewritten into a set of properties suited to be checked on an event-based TLM model. Secondly, an approach is defined to synthesize TLM properties into checkers to be adopted for dynamic ABV of the TLM model. In our work, we use a TL language to define TL assertions throughout a user defined TL-hierarchy.   



TL assertion language: In \cite{TL_ASS} an assertion specification language is presented which is based on formal definitions that allows the specification of TL properties and their execution in simulation. It derives the language from known ABV languages and extends these by the required TL functionality. It also explains how simulation traces of finite length can be checked against properties. Our work, on the other hand, is optimized for DFV. Nevertheless, the work in \cite{TL_ASS} was a great inspiration for our work, especially because it is based on industry experience. 


\section{Results}

The SoC design was written in PDVL. The number of synthesizable TRs and VTRs are listed in Tab \ref{tab_results} for each individual core and the complete SoC. MRPHS was used to compile the design into synthesizable SV code. The design and the VTRs are compiled into Gallina code. The compilation process of the SoC is completed within less than 10 seconds for each of the two outputs. For compilation and runtime evaluation, we use an i7, 2.6GHz CPU. 

The number of resulting theorems is listed in Tab. \ref{tab_results}. It also shows the complete consecutive execution time (CCET) of all proofs on a single thread for each individual core and the entire SoC. We introduce a maximal incremental runtime (MIRT), which is the worst case runtime to prove all relevant proofs based on an incremental change in the source code.

We developed a testbench in SV with the same coverage we achieve by the DFV flow. We use Verilator to run the regression suite. Tab \ref{tab_results} lists the simulation runtime (SRT) for each individual core and the entire SoC.

The execution time of a simulation based regression suite (SRT) is faster than the complete execution time of the DFV flow (CCET). This is mainly due to Coq theorem prover runtime issues. Nevertheless, DFV-based tests can be run in parallel to reduce this disadvantage, just like simulation-based regression suites. Alternatively, a faster theorem prover might eventually be used as our DFV is not limited to the Coq theorem prover.  

The runtime of the DFV flow when only an incremental update needs to be verified (MIRT), looks promising compared to the individual simulation runtime (SRT) of individual core related tests.

We see one of the most important advantages at the system level. While it is becoming increasingly difficult to write efficient system-level tests in simulation, we have found that it is very inviting to write top-level tests for our DFV flow. The possibility to reuse lower level proof is particularly helpful when solving software driver related issues. In this respect, we see a clear advantage of our demonstrated DFV flow over a simulation-based verification approach.

\section{Conclusion}
\label{conclusion}

We have demonstrated the use of deductive formal verification (DFV) to prove functional coverage and assertions as an alternative to simulation-based verification. One key contribution of DFV is the ability to symbolically verify complete coverage areas in a deductive manner. 
In order to benefit from verification reuse, we introduce a transaction level hierarchy, which enables the definition and verification of functional coverage and assertions from lower-level to system-level.

This hierarchical verification approach allows us to focus on level specific verification challenges and enables incremental updates when behavior only changes locally. It fills the gap that exists today between low-level SVA-based verification and the state-of-the-art coverage-based verification defined in UVM and PSS. At the same time, lower-level SW routines (e.g. peripheral drivers) become an integral part of the system verification at the same time.

\begin{table}[htb]
	\caption{Results for reference SoC design.}
	\begin{center}
		\begin{tabular}{|c|c|c|c|c|c|c|c|c|c|c|c|}
			\hline
			&TR&VTR&Theorems&Coq&Coq&Verilator\\
			\hline
			& & & & CCET & MIRT & SRT\\
			\hline
			& & &  & [sec] & [sec] & [sec]\\
			\hline
			RV32IMC & 107 & 144 & 219 & 52.4 & 5.01 & 27.8 \\
			\hline
			SDRAM & 93 & 160 & 197 & 55.9 & 3.49 & 27.0 \\
			\hline
			Ethernet & 72 & 89 & 106 & 18.3 & 2.48 & 14.0 \\
			\hline
			AES & 64 & 76 & 111 & 13.7 & 2.92 & 8.03 \\
			\hline
			SoC & 445 & 813 & 1048 & 186 & 6.26 & 125 \\
			\hline
		\end{tabular}
		\label{tab_results}
	\end{center}
\end{table}


\begin{thebibliography}{00}
	
	\bibitem{PDVL_spec} T. Strauch. PDVL Specification v0.1. [Online]. Available: https://github.com/cloudxcc/PDVL

	\bibitem{Gallina} Gallina Development Team. The Gallina specification language. [Online]. Available: https://coq.github.io/doc/v8.9/refman/language/gallina-specification-language.html

	\bibitem{Coq} Coq Development Team. The coq proof assistant. [Online]. Available: http://coq.inria.fr/

	\bibitem{SEC} Y. Morihiro, and T. Toneda, “Formal verification of Data-path Circuits Based on Symbolic Simulation”, Proc. of the Ninth Asian Test Symposium, 2000, 6th Dec, Taipei, Taiwan, pp. 1-8.

	\bibitem{PDVL} T. Strauch, "An Aspect and Transaction Oriented Programming, Design and Verification Language", IEEE Euromicro DSD 2017, 30 Aug. - 1 Sep., Vienna, Austria, pp. 30 - 39.

	\bibitem{PDVL_Gallina} T. Strauch, "Deductive Formal Verification of Synthesizable,
	Transaction-level Hardware Designs Using Coq", Design, Automation \& Test in Europe Conference \& Exhibition (DATE), 25-27 March 2024, Valencia, Spain, pp. 1-6.



	

	\bibitem{Syn_SVA} J. Long, and A. Seawright, "Synthesizing SVA Local Variables for Formal Verification", 44th ACM/IEEE Design Automation Conf (DAC),  4-8 June 2007, San Diego, CA, USA. pp. 75-80.





	\bibitem{Bluespec} R. Nikhil, ‘‘Bluespec SystemVerilog: Efficient, correct RTL from high level specifications,’’ Proc. 2nd ACM and IEEE Int. Conf. Formal Methods and Models for Co-Design, MEMOCODE’04, 23-25 June 2004, San Diego, CA, USA, pp. 69–70.
	
	\bibitem{BSV_MOD} M. Vijayaraghavan, N. Dave, and Arvind, “Modular Compilation of Guarded Atomic Actions”, ACM/IEEE Intern. Conf. on Formal Methods and Models for Codesign, MEMOCODE 2013, 18-20 Oct. 2013, Portland, OR, USA, pp. 177-188.

	\bibitem{KAMI} J. Choi, M. Vijayaraghavan, B. Sherman, A. Chlipala, and Arvind. “Kami: A Platform for High-Level Parametric Hardware Specification and Its Modular Verification”, Proceedings of the ACM on Programming Languages, Volume 1, Issue ICFP, Article No.: 24, pp 1–30.

	
	
	
	
	\bibitem{PSL_prover} K. Morin-Allory, M. Boulé, D. Borrione, and Zeljko Zilic, "Proving and Disproving Assertion Rewrite Rules with Automated Theorem Provers", IEEE Intern. High Level Design Validation and Test Workshop, 19-21 November 2008, Incline Village, NV, USA, pp. 56-63.
	
	\bibitem{RTL2TLM} Nicola Bombieri, Riccardo Filippozzi, Graziano Pravadelli and Francesco Stefanni, "RTL property abstraction for TLM assertion-based verification", Design, Automation \& Test in Europe Conference \& Exhibition (DATE), 9-13 March 2015, Grenoble, France, pp. 85-90.
	
	
	\bibitem{TL_ASS} W. Ecker, V. Esen, T. Steininger, M. Velten, and M. Hull, "Specification Language for Transaction Level Assertions", IEEE Intern. High Level Design Validation and Test Workshop, 8-10 November 2006, Monterey, CA, USA, pp. 77-84.

\end{thebibliography}
\end{document}